\documentclass[aps,pra,twocolumn,superscriptaddress]{revtex4-2}
\usepackage{graphicx}
\usepackage{dcolumn}
\usepackage{bm}
\usepackage{ulem}
\usepackage[T1]{fontenc}
\usepackage{physics}
\usepackage{braket}
\usepackage{amsmath}
\usepackage{mathrsfs}
\usepackage{svg}
\newcommand{\beq}{\begin{equation}}
\newcommand{\eeq}{\end{equation}}
\newcommand{\bei}{\begin{itemize}}			
\newcommand{\eei}{\end{itemize}}			

\usepackage{hyperref}
\hypersetup{
   pdfpagemode=None,
   pdfstartpage=1,
   pdfmenubar=true,
   pdftoolbar=true,
   colorlinks = true,
   linkcolor=blue,
   citecolor=blue,
   urlcolor=blue,
   bookmarksopen=false
 }

\begin{document}

\preprint{APS/123-QED}

\title{Electrically tunable liquid-crystal metasurfaces \\with patterned birefringence and dichroism}

\author{Paola Savarese}%
\thanks{These authors contributed equally to this work.}
\affiliation{%
Dipartimento di Fisica, Universit\`{a} degli Studi di Napoli Federico II, Complesso Universitario di Monte Sant'Angelo, Via Cintia, 80126 Napoli, Italy
}%
\author{Sarvesh Bansal}
\thanks{These authors contributed equally to this work.}
\affiliation{%
Dipartimento di Fisica, Universit\`{a} degli Studi di Napoli Federico II, Complesso Universitario di Monte Sant'Angelo, Via Cintia, 80126 Napoli, Italy
}%
\author{Maria Gorizia Ammendola}
\affiliation{%
Dipartimento di Fisica, Universit\`{a} degli Studi di Napoli Federico II, Complesso Universitario di Monte Sant'Angelo, Via Cintia, 80126 Napoli, Italy
}%
\affiliation{Scuola Superiore Meridionale, Via Mezzocannone, 4, 80138 Napoli, Italy}

\author{Raouf Barboza}
\affiliation{%
 Dipartimento di Scienze e Ingegneria della Materia, dell’Ambiente ed Urbanistica, Università Politecnica delle Marche,  Via Brecce Bianche, 60131 Ancona, Italy
}%
\author{Marcella Salvatore}
\affiliation{%
 Dipartimento di Fisica, Universit\`{a} degli Studi di Napoli Federico II, Complesso Universitario di Monte Sant'Angelo, Via Cintia, 80126 Napoli, Italy
}
\author{Stefano Luigi Oscurato}
\affiliation{%
 Dipartimento di Fisica, Universit\`{a} degli Studi di Napoli Federico II, Complesso Universitario di Monte Sant'Angelo, Via Cintia, 80126 Napoli, Italy
}
\author{Bruno Piccirillo}
\affiliation{%
 Dipartimento di Fisica, Universit\`{a} degli Studi di Napoli Federico II, Complesso Universitario di Monte Sant'Angelo, Via Cintia, 80126 Napoli, Italy
}%
\author{Francesco Di Colandrea}
\email{francesco.dicolandrea@unina.it}
\affiliation{%
 Dipartimento di Fisica, Universit\`{a} degli Studi di Napoli Federico II, Complesso Universitario di Monte Sant'Angelo, Via Cintia, 80126 Napoli, Italy
}%
\author{Lorenzo Marrucci}
\affiliation{%
 Dipartimento di Fisica, Universit\`{a} degli Studi di Napoli Federico II, Complesso Universitario di Monte Sant'Angelo, Via Cintia, 80126 Napoli, Italy
}%
\affiliation{CNR-ISASI, Institute of Applied Science and Intelligent Systems, Via Campi Flegrei 34, 80078 Pozzuoli (NA), Italy}
\author{Filippo Cardano}
\email{filippo.cardano2@unina.it}
\affiliation{%
Dipartimento di Fisica, Universit\`{a} degli Studi di Napoli Federico II, Complesso Universitario di Monte Sant'Angelo, Via Cintia, 80126 Napoli, Italy
}%


\begin{abstract}
Light propagation through artificially patterned anisotropic materials, such as dielectric metasurfaces, enables precise control of the spatio-vectorial properties of optical fields using highly transparent, thin, and flat optical elements. Liquid-crystal cells are a common realization of such devices. Optical losses are typically assumed to be polarization-independent and are therefore often overlooked in modeling these systems. In this work, we introduce electrically tunable liquid-crystal metasurfaces with patterned birefringence and dichroism, achieved by incorporating dichroic dye molecules into the liquid-crystal mixture. These dye molecules align with the liquid crystal, effectively coupling birefringence and dichroism effects. The behavior of these metasurfaces is described using non-unitary Jones matrices, validated through polarimetric measurements. In the case of devices that are patterned to form polarization gratings, we also characterize the diffraction efficiency as a function of the dichroism and birefringence parameters, which can be tuned jointly by applying an electric field across the cell. This study not only introduces a new class of optical components but also deepens our understanding of light propagation through anisotropic materials, where dichroism can naturally arise from bulk material properties or from reflection and transmission laws at their interfaces.
\end{abstract}

\maketitle

\section{Introduction}
Spin-orbit photonic technologies allow for accurate manipulation of spatio-vectorial features of optical fields by acting on optical polarization in a space-dependent manner~\cite{bliokh2015spin,cardano2015spin}. Their capability of generating structured light enables relevant applications, including particle trapping~\cite{trapping1,trapping2}, robust light guidance~\cite{waveguide_SOI,Slussarenko2016}, imaging~\cite{PhysRevLett.104.253601,PhysRevLett.91.233901,Bliokh:11}, and simulation of spin-dependent dynamics~\cite{qw_OAM}. Spin-orbit interactions of light also play a crucial role in topological photonics~\cite{RevModPhys.91.015006}, underlying the observation of unique phenomena, such as the subwavelength displacement of a circularly polarized beam at an interface, which is a typical manifestation of the spin-Hall effect of light~\cite{quantumspinhall,spinhallvortices,spinhallnearred}.

In the past decades, this functionality has been demonstrated for a variety of anisotropic materials, such as photonic crystals~\cite{PhysRevA.99.053845,Huang2024}, liquid crystals~\cite{ElKetara:12,doi:10.1126/science.aay4182,https://doi.org/10.1002/lpor.202400794}, epsilon-near-zero materials~\cite{Eismann:22}, as well as plasmonic~\cite{Karimi2014,Genevet:17,plasmonic_meta} and dielectric metasurfaces~\cite{doi:10.1073/pnas.1611740113,Ambrosio_2018,doi:10.1126/science.abi6860}. In this context, liquid-crystal metasurfaces (LCMSs), slabs of nematic liquid crystals (LC) artificially patterned on the micrometric scale~\cite{Kasyanova:18}, emerge as a versatile and compact solution to structure light polarization and beam transverse profile~\cite{Larocque_2016,Rubano:19}. These devices exploit gradients of the Pancharatnam-Berry phase to tailor optical beams across a wide spectral range~\cite{marruccipboe}. The prototypical example of LCMSs is the $q$-plate~\cite{marrucciq-plate}, wherein the liquid-crystal layer is aligned with an integer or semi-integer topological charge, which allows for partial or full conversion of spin angular momentum into orbital angular momentum. Since their first introduction, LCMSs have found applications in both classical and quantum domains~\cite{Marrucci_2011_review}, in particular microscopy~\cite{Yan:15}, communications~\cite{DAmbrosio2012,Sit:17}, metrology~\cite{DAmbrosio2013gear,Barboza2022}, and simulations~\cite{qw_OAM,DErrico:20,DiColandrea:23}.

In all experiments involving $q$-plates and their generalization to arbitrarily patterned devices, optical losses across the LC slab have been assumed to be polarization-independent, therefore no measurable effect was revealed, apart from a global attenuation of the incoming light intensity. Exceptions can be found in the context of LC polarization gratings, introduced specifically for polarization holography and polarization imaging~\cite{Todorov:84,pg1,Rubin2022}. In this paper, we go beyond this regime and extend the functionality of LCMSs to the case of polarization-dependent optical losses. In particular, these are introduced by doping the LC mixture with dichroic dye molecules, featuring differential absorption for two orthogonal polarization components. We show that the response of this new class of LCMSs can be effectively modeled as a non-unitary Jones matrix, where the distance from the unitary case is proportional to the dichroic power of the cell. A similar formalism has also been adopted in a recent study, introducing thermally tunable dichroic LCMSs~\cite{Suzuki:24}. Our devices are electrically tunable, that is, when an electric field is applied across the cell, a torque on LC molecules tilts them out of the slab plane, modifying both birefringence and dichroism. This model is experimentally validated on a few samples of dichroic LCMSs, both in the case of uniform and spatially patterned optic-axis orientations. In the latter scenario, we specifically focus on polarization gratings, given their use in several fields, and provide a characterization of their diffraction efficiency as a function of the dichroic parameter. 



\begin{figure}[ht]
  \centering
\includegraphics[width=0.5\textwidth]{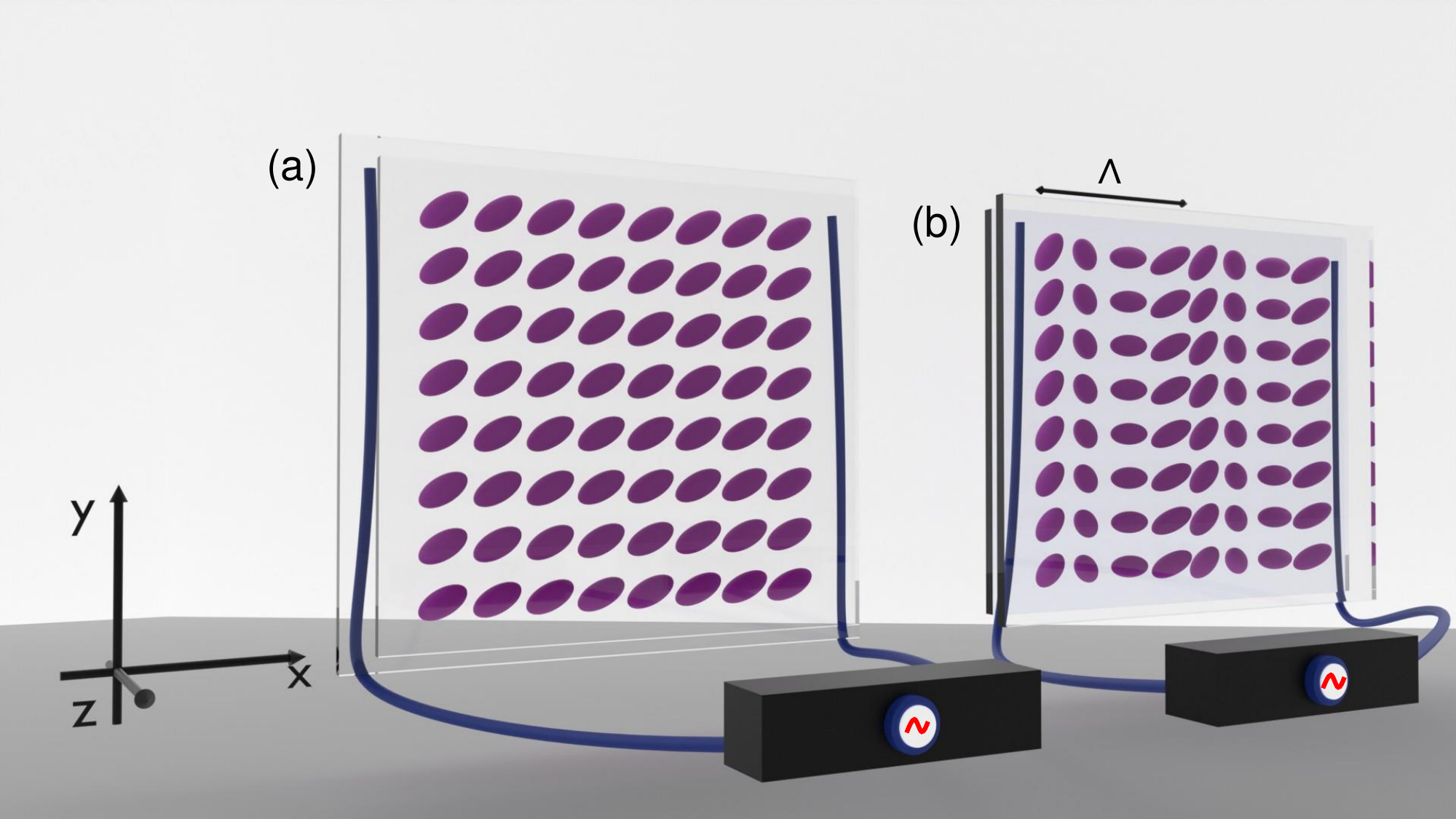}
  \caption{\textbf{Dichroic LCMSs.}~(a) Homogenous LCMS and (b)~inhomogeneous LCMS, with the latter exhibiting a periodic pattern along the $x$ axis.}
  \label{fig:plates}
\end{figure}

\section{Dichroic LCMS{$\text{s}$}} 
LCMSs can be effectively modeled as slabs of birefringent uniaxial materials~\cite{yeh2009optics}. Their action is typically described through the Jones formalism, where the field is described by a 2D complex vector. The vector entries represent the complex amplitudes of the field associated with two orthogonal polarization states, serving as a basis. If diffraction within the sample can be neglected, the output field of the LCMS is obtained by applying the device's Jones matrix to the input Jones vector. To begin, consider a LCMS with a uniform pattern. In the basis of the extraordinary and ordinary axes, the Jones matrix for such a system is expressed as
\begin{equation}
\mathcal{L}=\begin{pmatrix}
e^{i k n_e d} && 0\\
0 && e^{i k n_o d}
\end{pmatrix},
\label{eqn:jones1}
\end{equation}
where ${k=2\pi/\lambda}$ is the incident light wavevector, with ${\lambda}$ the optical wavelength, $n_e$ ($n_o$) is the refractive index along the extraordinary (ordinary) axis, and ${d}$ is the sample thickness. For simplicity, we can assume that the extraordinary (ordinary) axis coincides with the $x$ ($y$) direction, and photons propagate along ${z}$. This model is compatible with polarization-independent losses, which affect the propagation of two polarization components in the same way, and can therefore be neglected. 

We aim to introduce linear dichroism in our device, which corresponds to differential absorption depending on the input linear polarization. In LC displays, a typical solution relies on the so-called \textit{guest-host} effect~\cite{guesthost_top}, wherein a guest dichroic compound, such as an absorbing dye, is incorporated into the host LC matrix. In our experiment, the LC mixture is doped with organic dye molecules, specifically {1,4-Diaminoanthraquinone}. The Jones matrix of Eq.~\eqref{eqn:jones1} is thus generalized as
\begin{equation}
\mathcal{L}=\begin{pmatrix}
e^{i k n_e d}e^{-\alpha_e d} && 0\\
0 && e^{i k n_o d}e^{-\alpha_o d}
\end{pmatrix},
\label{eqn:jones2}
\end{equation}
where ${\alpha_e}$ (${\alpha_o}$) is the absorption coefficient along the extraordinary (ordinary) axis, typically with ${\alpha_e>\alpha_o}$~\cite{marrucci2000role}. The matrix of Eq.~\eqref{eqn:jones2} is indeed a non-unitary Jones matrix, whose birefringence and dichroism can be formally combined as follows:
\begin{equation}
\mathcal{L}=e^{i\bar \zeta/2}\begin{pmatrix}
e^{i \zeta/2} && 0\\
0 && e^{-i \zeta/2}
\end{pmatrix},
\label{eqn:jones3}
\end{equation}
where $\zeta=\delta+i\eta$ and $\bar\zeta=\bar\delta+i\bar\eta$ are complex parameters, with 
 \begin{equation}
\begin{split}
    \delta&=kd\left({n_e-n_o}\right), \\
    \eta&=d\left(\alpha_e-\alpha_0\right), \\
     \bar\delta&=kd\left({n_e+n_o}\right), \\
    \bar\eta&=d\left(\alpha_e+\alpha_o\right).
\end{split}
\label{eqn:definition}
\end{equation}
In Eq.~\eqref{eqn:jones3}, $e^{i\bar\zeta/2}$ accounts for a global phase shift and attenuation factor. 

If the LC molecules lie at an angle $\theta$ in the $xy$ plane, the dye molecules align with the underlying birefringent substrate~\cite{guesthost_top}. Accordingly, the Jones matrix of Eq.~\eqref{eqn:jones3} is transformed as
\begin{equation}
\mathcal{L}_\theta=e^{i\bar \zeta/2}\begin{pmatrix}
\cos{\frac{\zeta}{2}}+i\sin{\frac{\zeta}{2}\cos{2\theta}}&& i\sin{\frac{\zeta}{2}}\sin{2\theta}\\
i\sin{\frac{\zeta}{2}}\sin{2\theta} && \cos{\frac{\zeta}{2}}-i\sin{\frac{\zeta}{2}\cos{2\theta}}
\end{pmatrix}.
\end{equation}
The latter assumes a simpler expression in the circular polarization basis, where ${\ket{L}=(1,0)^T}$ (${\ket{R}=(0,1)^T}$) is the left-handed (right-handed) circular polarization state:
\begin{equation}
\mathcal{L}_\theta=e^{i\bar \zeta/2}\begin{pmatrix}
 \cos{\frac{\zeta}{2}}&i\sin{\frac{\zeta}{2}e^{-2i\theta}}\\
i\sin{\frac{\zeta}{2}e^{2i\theta}}&\cos{\frac{\zeta}{2}}
\end{pmatrix}.
\label{eqn:jonescircular}
\end{equation}
When $\theta$ is uniform across the transverse plane, we refer to these devices as homogeneous LCMSs (see Fig.~\ref{fig:plates}(a)). However, LC molecules can be photoaligned according to a specific pattern ${\theta(x,y)}$, as shown in Fig.~\ref{fig:plates}(b). 
In the following, we provide a characterization of dichroic LCMSs in both configurations. 

\subsection{Dichroic homogeneous LCMSs}
\label{sec:homogeneous}
\begin{figure}[b]
 \includegraphics[scale=0.40]{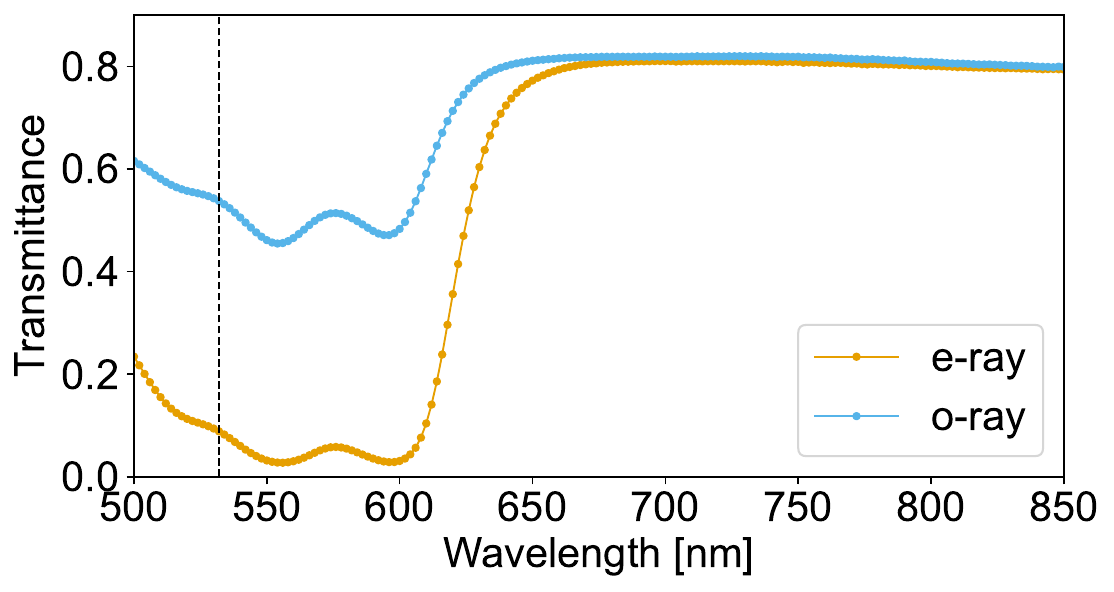}
  \caption{\textbf{Spectral characterization of dichroic LCMSs.} Measured transmittance of extraordinary and ordinary components for different wavelengths. A vertical bar at ${\lambda=532}$~nm indicates the conditions of our experiment.}
  \label{fig6}
\end{figure}
The dichroic coefficient ${\eta}$ is wavelength-dependent. To characterize this spectral response, absorption measurements are carried out for homogeneous samples. From the results reported in Fig.~\ref{fig6}, it is clear that the dichroic power of the LCMSs significantly decreases above 630~nm, where the absorption of extraordinary and ordinary components is comparable. Accordingly, ${\lambda=532}$~nm is selected as the operating wavelength to maximize the dichroism of our devices while aligning with the availability of laser sources in our laboratory.
\begin{figure}[h!]
\includegraphics[scale=0.40]{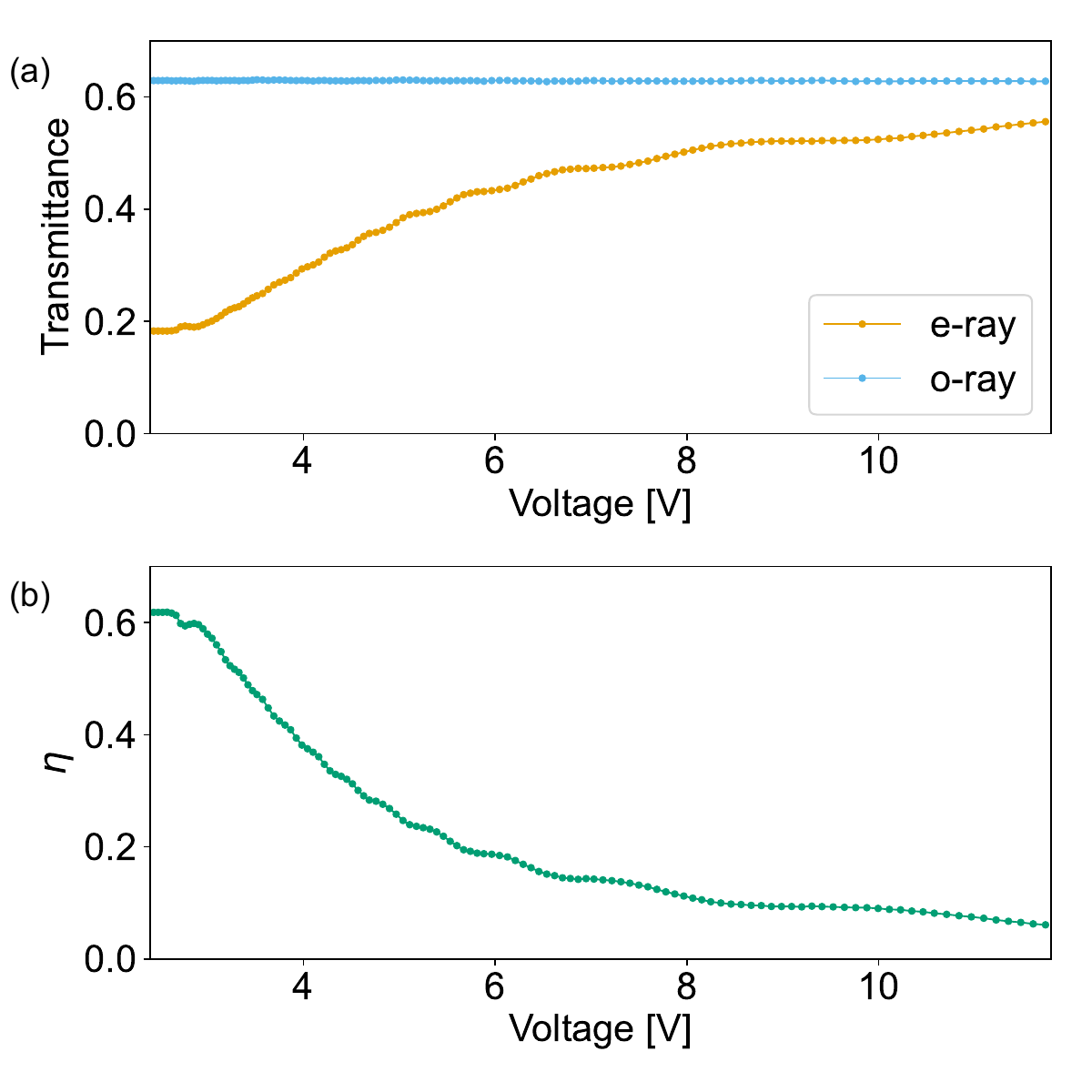}
  \caption{\textbf{Linear dichroism as a function of voltage.} (a)~Measured transmittance of extraordinary and ordinary components at ${\lambda=532}$~nm for different input voltages. (b)~Extracted dichroic parameter ${\eta}$ from the recorded transmittance. At high voltages, the dichroic power of the cells vanishes.}
  \label{homogeneous_waveplate_linear}
\end{figure}
The application of an external electric field along $z$ causes the LC molecules to tilt out of plane toward the $z$-axis (see Fig.~\ref{fig:plates}), altering the extraordinary refractive index of the LC layer. As the field strength increases, this index approaches the value of the ordinary refractive index. In conventional LCMSs, this effect is used to tune the effective birefringence parameter ${\delta}$~\cite{Piccirillo2010,Rubano:19}. Being the orientation of dye molecules coupled to that of LC, in the case of dichroic LCMSs the dichroic power of the cell changes as well when adjusting the external electric field. We characterized this effect by recording the transmitted power for the extraordinary and ordinary waves at different values of the applied voltage. Specifically, an oscillating square signal at approximately 4 KHz is used. From the transmittance curves shown in Fig.~\ref{homogeneous_waveplate_linear}(a), the dichroic parameter ${\eta}$ is extracted at each voltage (cf.~Eqs.~\eqref{eqn:jones3}-\eqref{eqn:definition}):
\begin{equation}
\eta=\frac{1}{2}\log{\frac{P_\text{o}}{P_\text{e}}},
\end{equation}
where ${P_\text{o}}$ (${P_\text{e}}$) is the transmitted power of the ordinary (extraordinary) component.  
As expected, at high voltages the LC molecules are approximately parallel to the propagation direction, which results in a significant reduction of dichroism, as shown in Fig.~\ref{homogeneous_waveplate_linear}(b). In all figures showing the response of these devices to the applied electric field, the reported voltage values represent peak-to-peak amplitudes.
\begin{figure}[h!]
\includegraphics[scale=0.43]{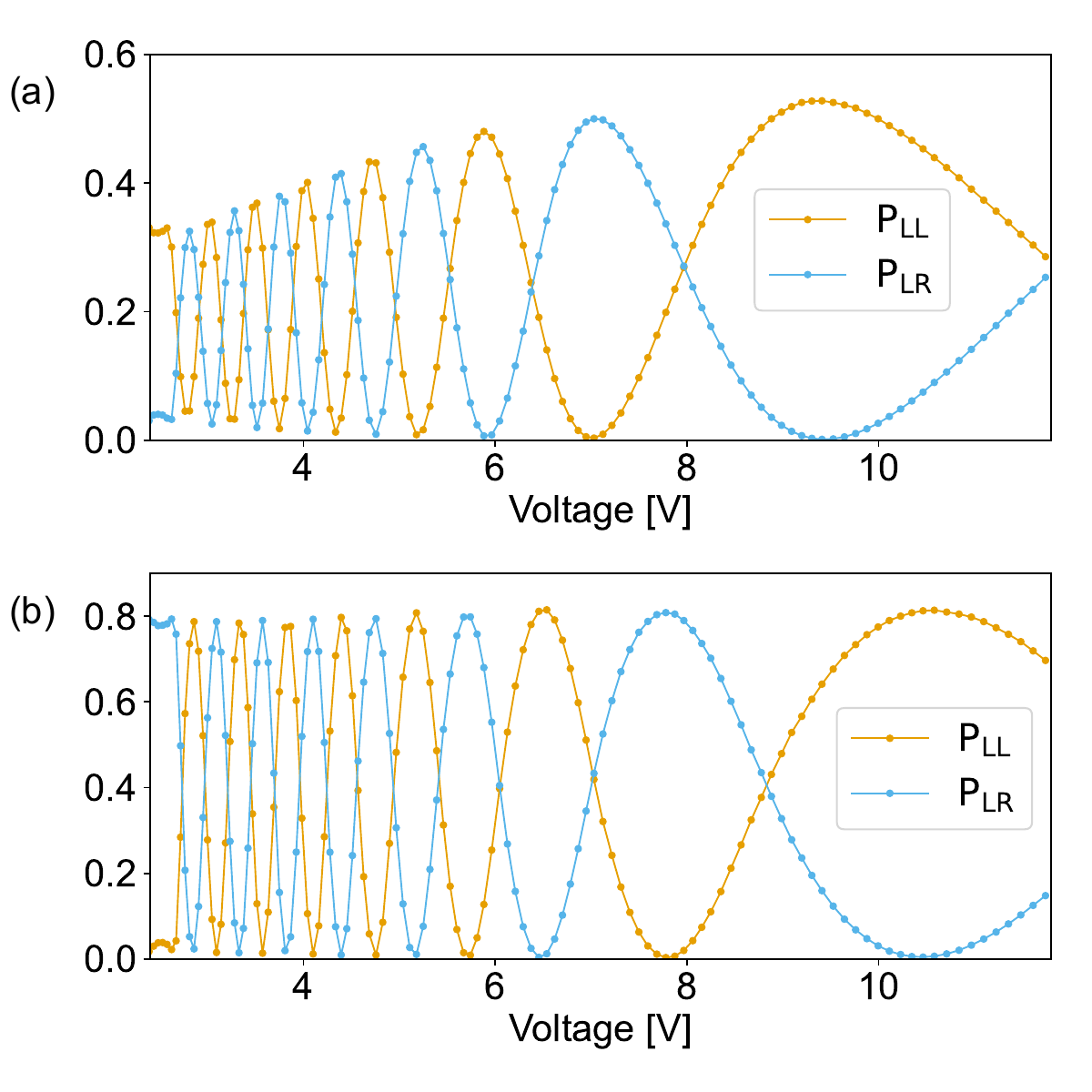}
  \caption{\textbf{Characterization of dichroic LCMSs with circularly polarized light.} Measured power (normalized to the total input power) of the converted ($P_\text{LR}$) and unconverted ($P_\text{LL}$) circular polarization components from (a) a dichroic and (b) a non-dichroic LCMS. The varying ratio between maxima and minima is a signature of the dichroic character of the cell.}
  \label{homogeneous_waveplate_circular}
\end{figure}

Homogeneous LCMSs are also characterized with circularly polarized light. In particular, we illuminate the sample with a left-handed input beam, and project the transmitted light on the two orthogonal circular polarization components. From Eq.~\eqref{eqn:jonescircular}, we obtain
\begin{equation}
\begin{split}
P_{\text{LR}}&=e^{-\bar\eta}\left|\sin{\frac{\zeta}{2}}\right|^2=e^{-\bar\eta}\frac{\cosh{\eta}-\cos{\delta}}{2},\\
P_{\text{LL}}&=e^{-\bar\eta}\left|\cos{\frac{\zeta}{2}}\right|^2=e^{-\bar\eta}\frac{\cosh{\eta}+\cos{\delta}}{2},
\end{split}
\label{eqn:convunconv}
\end{equation}
where ${P_\text{LR}}$ (${P_\text{LL}}$) is the power of the right (left) projection, corresponding to the converted (unconverted) fraction of incoming light. In the limit ${\eta\rightarrow0}$, up to the attenuation factor $e^{-\bar\eta}$, the functionality of standard LCMSs is recovered, with the unitary condition expressed as
\begin{equation}
P_{\text{LR},\eta\rightarrow0}+P_{\text{LL},\eta\rightarrow0}=\sin^2{\frac{\delta}{2}}+\cos^2{\frac{\delta}{2}}=1.
\end{equation}
The joint presence of birefringence and dichroism, and their dependence on the external field, can be revealed by measuring the transmitted power of the converted and unconverted light at different voltages. In particular, the relative amplitude of the minima and maxima of the two components does not stay constant (see Fig.~\ref{homogeneous_waveplate_circular}(a)), as both the birefringence ${\delta}$ and the dichroic parameter ${\eta}$ depend on the applied voltage, unlike non-dichroic LCMSs (see Fig.~\ref{homogeneous_waveplate_circular}(b)). Furthermore, minima of both $P_\text{LR}$ and $P_\text{LL}$, that are expected to vanish in the unitary case, take finite values due to the presence of dichroism. Nevertheless, as in the case of standard LCMSs, the maxima (minima) of the converted (unconverted) component correspond approximately to an effective birefringence equal to half-wave retardation (see Eq.~\eqref{eqn:convunconv}): ${\delta=(2n+1)\pi}$, with ${n}$ an integer number. Vice versa, the minima (maxima) of the converted (unconverted) component correspond to full-wave retardation: ${\delta=2n\pi}$. Quarter-wave retardation (${\delta=(2n+1)\pi/2}$) is obtained when ${P_\text{LR}=I_\text{LL}}$.

\subsection{Dichroic patterned LCMSs}
As discussed above, LC molecules can be photoaligned so as to match a specific pattern. Here, we focus on linear polarization gratings \cite{Todorov:84}, where the optic-axis orientation features a 1D linear modulation, ${\theta(x)=\pi x/\Lambda}$. $\Lambda$ is a characteristic distance, set to $5$~mm in our experiment, representing the spatial period of the grating. Originally introduced in the context of polarization holography~\cite{Todorov:84}, they have recently been used for the simulation of 2D quantum walks in the light transverse-momentum space~\cite{DErrico:20}, where they have been dubbed $g$-plates~\cite{DErrico:20}. Such devices have also been recently employed for enhanced metrological protocols~\cite{Barboza2022}.

\begin{figure}[h!]
\includegraphics[scale=0.9]{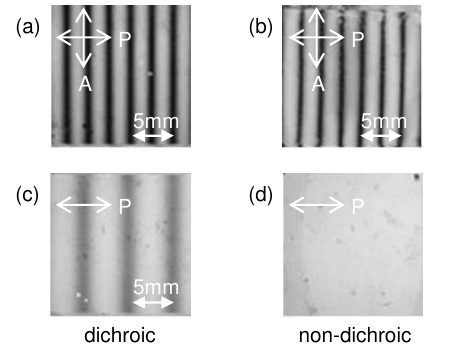}
\caption{\textbf{LC polarization gratings.} (a)~A dichroic and (b) a non-dichroic LCMS, patterned with a grating structure, are observed between crossed polarizers P and A. When removing the analyzer A, (c) the grating pattern is clearly visible in the dichroic device, while (d) it is absent for standard LCMSs.}
  \label{grating_image}
\end{figure}

Polarization gratings, or $g$-plates, act by deflecting incident light in opposite directions depending on the input polarization~\cite{Todorov:84,Gou:17,Sakamoto:21}. Figure~\ref{grating_image}(a) shows the pattern emerging when observing a dichroic $g$-plate between crossed polarizers. This is very similar to the typical fringe pattern of a non-dichroic cell, having bright and dark fringes alternating with a period $\Lambda/2$, as illustrated in Fig.~\ref{grating_image}(b). However, the dichroic character of the plate can be revealed by removing the analyzer (A) while maintaining the polarizer (P). In fact, the differential absorption of light by dye molecules reveals the grating structure even without an analyzer (see Fig.~\ref{grating_image}(c)-(d)). In this case, the spatial period of the fringes is $\Lambda$. 

\begin{figure*}
\includegraphics[scale=0.35]{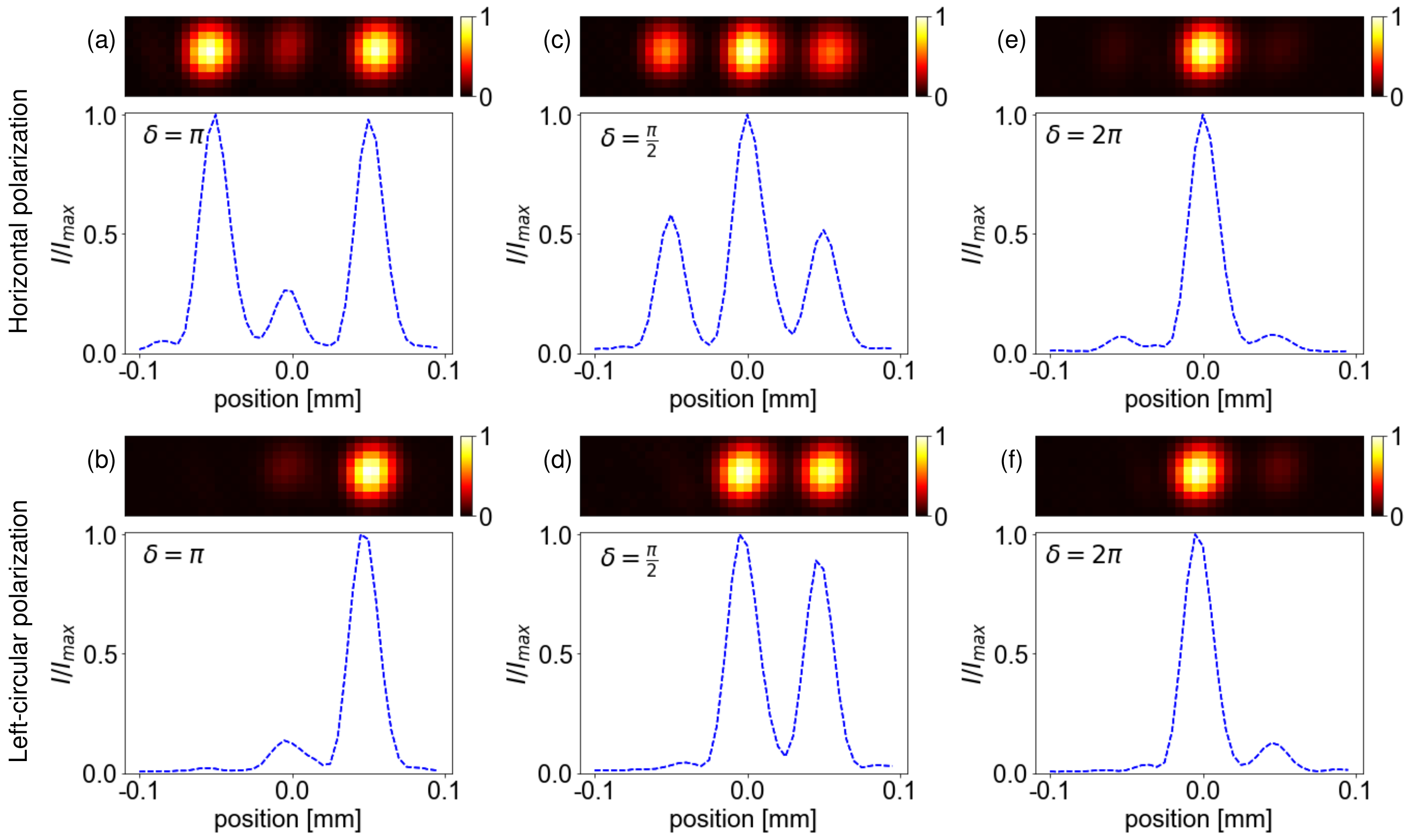}
    \caption{\textbf{Diffraction from a dichroic $g$-plate.} Measured far-field distributions and corresponding line intensity profiles from a dichroic $g$-plate operating at half-wave (a-b), quarter-wave (c-d), and full-wave (e-f) retardation, for horizontally (top) and left-circular (bottom) polarized input states. Retardations corresponding to $\delta=\pi$, $\frac{\pi}{2}$, and $2\pi$ are obtained by applying a peak-to-peak voltage equal to 2.82 V, 2.95 V, and 3.05 V, respectively.}
    \label{fig3}
\end{figure*}

The $g$-plate device modifies the input beam by adding a quantized amount of transverse momentum ${\pm\Delta k_\perp}$, given by ${\Delta k_\perp=2\pi/\Lambda}$, depending on the input polarization (cf.~Eq.~\eqref{eqn:jonescircular}). In other words, it creates optical modes characterized by a wavefront slightly tilted in the $xz$ plane. These modes carrying discrete units of transverse momentum can be sorted on a CCD camera placed in the focal plane of a converging lens, as long as ${w_0\geq\Lambda}$, with $w_0$ the input beam waist. In this scenario, the modes appear as separate Gaussian spots arranged on a line~\cite{DErrico:20}. Representative far-field distributions recorded on the camera are shown in the top insets of Fig.~\ref{fig3}. 

In the following, we provide a characterization of the diffraction efficiency of a dichroic $g$-plate for linear and circularly polarized input states and for different values of the birefringence parameter, selected as illustrated in Section \ref{sec:homogeneous}. When considering a horizontally polarized input beam, ${\ket{H}=\left(\ket{L}+\ket{R}\right)/\sqrt{2}}$, an ideal non-dichroic $g$-plate at half-wave retardation can be used to perfectly sort optical modes being left or right polarized, by adding a single unit of transverse momentum ${\Delta k_\perp}$ with opposite signs. This is not the case for a dichroic plate, where the dichroism is responsible for a residual fraction of incoming light sent to the zero order, as shown in Fig.~\ref{fig3}(a). A similar behavior is observed for a left-circular input, wherein only the first diffraction order is expected in the non-dichroic case (see Fig.~\ref{fig3}(b)). In the case of quarter-wave plate retardation, the dichroic character manifests itself as a global attenuation equally affecting the diffracted and unmodulated light, as shown in Fig.~\ref{fig3}(c)-(d). An interesting case is provided by full-wave retardation, where non-dichroic $g$-plates act as the identity operator, while the presence of dichroism in our samples drives a small portion of the input light to the first diffraction orders (see Fig.~\ref{fig3}(e)-(f))

\begin{figure}[h!]
\includegraphics[scale=0.45]{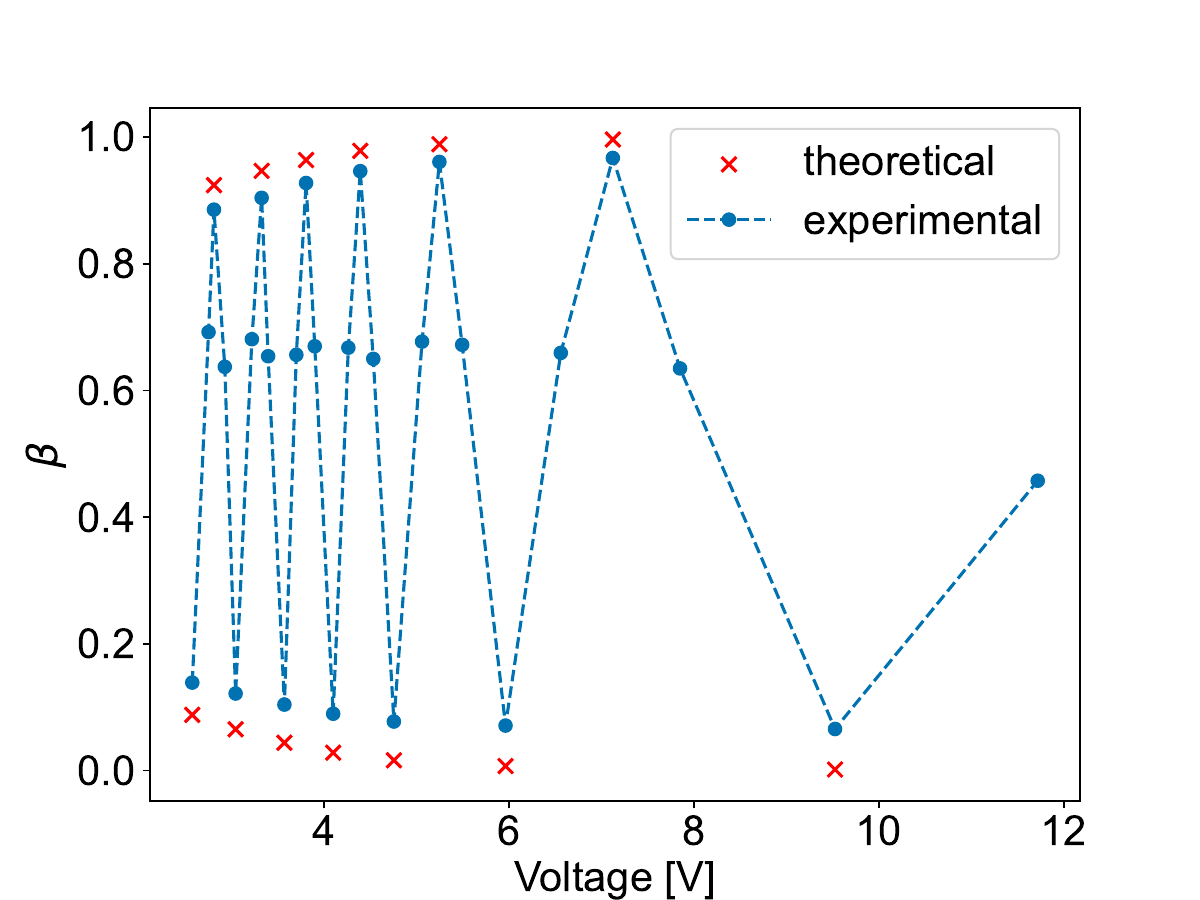}
  \caption{\textbf{Diffraction efficiency of a dichroic $g$-plate.} Measured values of diffraction efficiency of a dichroic $g$-plate at different voltages, obtained when shining the cell with circularly polarized light. As predicted from Eq.~\eqref{eqn:convunconv}, the maxima and minima of the diffraction efficiency do not stay constant, as at larger values of the applied voltage the dichroic power of the cell is reduced.}
  \label{fig:beta}
\end{figure}

By collecting intensity distributions in the far field, as those illustrated in Fig.~\ref{fig3}, the diffraction efficiency ${\beta}$ of the dichroic $g$-plate at several values of the applied voltage is extracted. Specifically, we compute as $P_{-1}, P_{0}, P_{1}$ the integrated intensity of the corresponding diffraction orders. $\beta$ is then obtained as
\begin{equation}\label{eq:beta}
\beta=\frac{P_{-1}+P_{1}}{P_{-1}+P_{1}+P_{0}}.
\end{equation}
Essentially, it measures the ratio between correctly diffracted light and total transmitted light, assuming that there is no diffraction into higher-order modes, which is the case for linear phase gradients. Left-circular polarization is used as input. As expected, maxima (minima) of $\beta$ appear in correspondence with half-wave (full-wave) retardation. However, unlike non-dichroic devices, the diffraction efficiency does not take the same values at different maxima and minima, as increasing the voltage reduces the dichroic power of the cell (cf.~Fig.~\ref{homogeneous_waveplate_linear}). Experimental results are reported in Fig.~\ref{fig:beta}. They are compared with theoretical estimates, extracted from Eq.~\eqref{eqn:convunconv} with values of $\eta$ reported in Fig.~\ref{homogeneous_waveplate_linear}(b). 

\section{Conclusion}

We introduced a new class of LCMSs that exhibit polarization-dependent losses. The optical behavior of these devices can be described by a non-unitary Jones matrix, formally equivalent to that of a standard waveplate with a complex birefringence parameter. This model has been experimentally validated on both homogeneous and structured metasurfaces, where the spin-orbit functionality embeds the dichroic feature. \\
We demonstrated that the application of an external voltage serves as a convenient control for tuning the dichroic character of the cell, with the only limitation being that the birefringence and dichroic properties are inherently coupled and cannot be adjusted independently.\\
The new devices hold great promise for the simulation of non-unitary dynamics and non-Hermitian topological phases~\cite{PhysRevLett.119.130501_2,gong2018topological,RevFlore,Nasari:23}.

\section*{Methods}
\subsection*{Fabrication}
Two glass plates, pre-coated with a thin film of ITO (Indium-Tin-Oxide, a transparent conductive material), undergo two-stage cleaning in an ultra-sonication chamber, the first with a 5\% solution of phosphate-free alkaline detergent and distilled water, and the second with distilled water only, both for 60 mins at ${60^\circ}$. The plates are then dried in the oven for 60 mins at $100^\circ$. This is followed by a 45-min Ozone-UV exposure, facilitating the subsequent azo-dye deposition. Glass plates are spin-coated with a thin film of photosensitive azo-dye. After spin-coating, the plates are sandwiched together with the help of $17~\mu$m silica spacers. The desired pattern is imprinted using a well-established photoalignment technique~\cite{Rubano:19}, allowing for the azo-dye molecules to be oriented point by point by rotating the linear polarization of an incident beam at $445$~nm. After photoalignment, the cell is eventually filled with a mixture of nematic liquid crystals saturated with 1,4-Diaminoanthraquinone, which penetrates by capillarity the sample pre-heated at $100^\circ$.\\ \\
\noindent \textbf{Acknowledgments.}
This work was supported by the PNRR MUR project PE0000023-NQSTI.
\newline \noindent \textbf{Disclosures.}
The authors declare no conflicts of interest.
\newpage
\bibliography{main}

\end{document}